\documentclass[runningheads,a4paper]{llncs}

\usepackage{amssymb}
\usepackage{amsmath}
\usepackage{graphicx}
\usepackage{color}

\usepackage[export]{adjustbox}  

\usepackage{colortbl}
\usepackage{pstricks,pst-grad}
\usepackage{pst-pdf}
\newrgbcolor{veryLightGrey}{0.96 0.96 0.96}

\usepackage{xcolor} 
\usepackage[urlcolor=blue]{hyperref}      % links to citations
\hypersetup{
     colorlinks = true,                    % text and not border
     citecolor = {blue},
     linkcolor = {violet},
           }
\urldef{\mailsa}\path|{kubandt,gros07}@itp.uni-frankfurt.de|

\begin{document}

\mainmatter  % start of an individual contribution

\title{Embodied robots driven by self-organized 
environmental feedback}
\titlerunning{Self-organized embodied robots}

\author{Frederike Kubandt$^1$, Michael Nowak$^1$, Tim Koglin$^1$, \\ 
        Claudius Gros$^1$, Bulcs\'u S\'andor$^{2,1}$}
\authorrunning{Train of cars} % for left-page running title
\institute{$^1$Institute for Theoretical Physics, Goethe University Frankfurt, Germany\\
           $^2$Department of Physics, Babes-Bolyai University,
Cluj-Napoca, Romania\\
\mailsa\\
\url{http://itp.uni-frankfurt.de/~gros}}

\maketitle

\begin{abstract}
Which kind of complex behavior may arise from self-organizing principles? We
investigate this question for the case of snake-like robots composed of
passively coupled segments, with every segment containing two wheels actuated
separately by a single neuron. The robot is self organized both on the level of
the individual wheels and with respect to inter-wheel coordination, which
arises exclusively from the mechanical coupling of the individual wheels and
segments. For the individual wheel, the generating principle proposed results
in locomotive states that correspond to self-organized limit cycles of the
sensorimotor loop. 

Our robot interacts with the environment by monitoring the state of its
actuators, that is via propriosensation. External sensors are absent. In a
structured environment the robot shows complex emergent behavior that includes
pushing movable blocks around, reversing direction when hitting a wall and
turning when climbing a slope. On flat grounds the robot wiggles in a
snake-like manner, when moving at higher velocities. We also investigate the
emergence of motor primitives, viz the route to locomotion, which is
characterized by a series of local and global bifurcations in terms of
dynamical system theory.  \end{abstract}

%%%%%%%%%%%%%%%%%%%%%%%%%%%%%%%%%%%%%%%%%%%%%%%%%%%%%%%%%%%%%%%%%%%%%%55
%%%%%%%%%%%%%%%%%%%%%%%%%%%%%%%%%%%%%%%%%%%%%%%%%%%%%%%%%%%%%%%%%%%%%%55

%%%%%%%%%%%%%%%%%%%%%%%%%%%%%%%%%%%
\section{Introduction}
%%%%%%%%%%%%%%%%%%%%%%%%%%%%%%%%%%%

Wheeled snake-like robots \cite{tanaka2015control}
are a class of hypermobile robots \cite{granosik2014hypermobile}
that are able to navigate flexibly through rough terrains 
and restricted geometries. Movements may be generated either
via central pattern generators \cite{hopkins2009survey}, or
via top-down commands \cite{pfotzer2017autonomous}, with the
latter being a challenging task when a large number of actuators 
is involved. An alternative is autonomous decentralized
control, which has been studied for the case of serpentine robots 
in terms of a chain of locally coupled oscillators 
\cite{sato2011applicability,sato2011decentralized}, and 
neurally-inspired generating schemes able of
sensorless pathfinding \cite{boyle2013adaptive}.

A key rationale for developing biologically inspired robots 
is the drive for robust and highly adaptive designs
\cite{hirose2004biologically,liljeback2012review}. Similarly,
this is most of the time also the motive for studying how
adaptive locomotion can be realized \cite{aoi2017adaptive},
f.i.\ with soft robots \cite{calisti2017fundamentals}.
Abstracting from the direct engineering benefit, it is 
of particular interest to study generating mechanisms of 
locomotion in general, an approach taken here. Our focus
is on compliant locomotion generated by self-organizing 
dynamical systems, which may take the
form of either limit-cycle \cite{martin2016closed}
or playful behavior \cite{der2006rocking}. 

Compliance, which denotes the ability of a robot to 
react elastically to environmental feedback \cite{sprowitz2013towards},
may be achieved in several distinct ways, which include
from the engineering perspective suitably designed actuators \cite{van2009compliant} 
and control algorithms \cite{calanca2016review}. Compliant 
behavior can emerge on the other side also through the 
reciprocal dynamical coupling of control, body and environment
\cite{pfeifer2012challenges}, the sensorimotor loop. A particularly
interesting limit is here, from the perspective of complex system
theory, the limit of a fully reactive and hence embodied 
controller. In this limit, the controller is inactive in 
absence of environmental feedback, with the consequence that
the sensorimotor feedback has not only a modulating effect on
locomotion, becoming instead essential. Locomotion then arises 
via limit cycles and chaotic attractors that emerge within the 
sensorimotor loop, the telltale sign of self-organized locomotion. 
It is hence important to ask, as we will do in this study, how 
locomotion is generated in terms of a dynamical systems bifurcation 
diagram.

%----------------------------------------------------------
\begin{figure}[!t]
\centering
\includegraphics[height=0.5\textwidth]{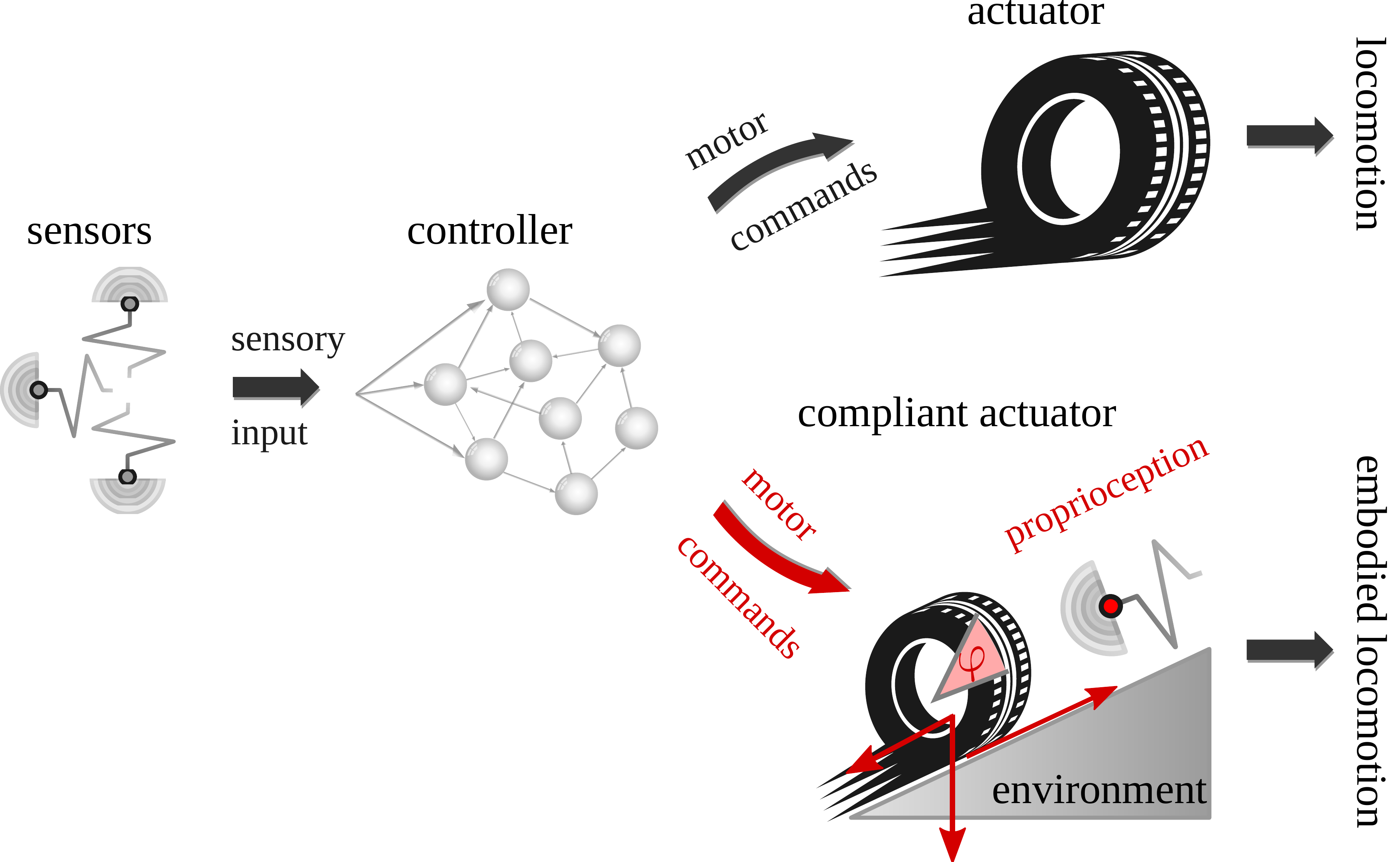}
\caption{Illustration of possible control schemes. Sensory
information is processed, e.g.\ by a neural network, and motor 
commands sent to the actuators. The actuators may respond either 
rigidly (top), or elastically, viz compliant (bottom).
Compliant actuators may be realized, as illustrated here, via 
a direct feedback loop involving the state of the actuator 
(propriosensation). In the limiting case of an embodied
actuator, as considered in this study, locomotion occurs also in
the absence of a modulatory top-down signal.
}
\label{fig:control_topDownCompliant}
\end{figure}
%----------------------------------------------------------

Decomposing complex behavior into a series (or into a 
superposition) of basic reusable building blocks, the
motor primitives \cite{flash2005motor}, is a well studied 
approach for reducing the control problem of complex robots. 
Movement primitives may be modeled by nonlinear dynamical 
systems \cite{ijspeert2002movement} using, e.g., 
Gaussian mixture models \cite{khansari2011learning},
where the parameters of the dynamical system are either
uniquely defined or drawn from a suitable distribution
\cite{paraschos2013probabilistic,amor2014interaction}.
Motor primitives can emerge also from embodied dynamics
in terms of chaotic itinerancy \cite{park2017chaotic},
or, alternatively, as self-organized attracting states 
in the sensorimotor loop \cite{sandor2015sensorimotor}, 
that is within the state space comprising the controller, 
the body of the robot and the environment. Here we propose 
a new type of self-organized controller for wheeled robots
that leads to multiple fixpoint and limit-cycle 
attractor states and hence to self-organized motor 
primitives in the sensorimotor loop. With the
behavior of the robot being self organized on the 
level of the individual wheels and with respect to 
inter-wheel coordination, the resulting dynamics reflects 
its affordances \cite{chemero2007gibsonian}
when placed in simple but structured environments.

% ------------
\subsection{Control frameworks and the sensorimotor loop}
% ------------

Several in part non-exclusive routes for the 
generation of locomotion in robots and animats 
do exist in generic terms. Standard
top-down control, as illustrated
in Fig.~\ref{fig:control_topDownCompliant},
consists of a central processor 
generating motor commands either reactively,
in response to sensory inputs, or deliberately
on its own \cite{nakhaeinia2011review}. The 
actuator may be in turn stiff, as for industrial
robots, or compliant, as for the muscles and tendons
of animals, reacting either passively or actively 
to external forces \cite{van2009compliant}. For
the latter case, as sketched in
 Fig.~\ref{fig:control_topDownCompliant}, the
actuator changes its stiffness upon sensing 
its own state. Compliance arises then
in response to propriosensation.

We are interested in locomotion that arises through 
the interaction of the degrees of freedom of the robot, 
including both internal variables and the body, with 
environmental feedback. The combined variables of the 
resulting sensorimotor loop constitute then the phase 
space for dynamical attracting states, fixed points, 
limit cycles and chaotic attractors, that correspond 
to self-organized behavioral primitives. The locomotion 
generated in this way is highly compliant in the sense 
that the attracting states in the sensorimotor loop 
respond elastically to additional top-down commands 
changing internal parameters.

%%%%%%%%%%%%%%%%%%%%%%%%%%%%%%%%%%%
\section{Locomotive principles}
%%%%%%%%%%%%%%%%%%%%%%%%%%%%%%%%%%%

Studies of real-world and simulated robots may focus 
either on performance, and its improvement, or on the 
generative capabilities of locomotive principles.
The latter approach is gaining in importance in view of
a recent study of the neural coding of leg dynamics 
in flies, which showed that the dynamics of the 
leg becomes dysfunction once the feedback loop between 
leg proprioception and motor commands is cut
\cite{mamiya2018neural}. These findings imply that
self-organization plays a commanding role in fly
locomotion. Distributed computations has been found
to be of relevance for the nematode C.~elegans
\cite{kaplan2018sensorimotor}. Here we concentrate 
on generative principles that are time reversal 
symmetric in the sense that a given set of internal 
parameters allows the robot to move both forwards and 
backwards. The direction selected by the robot then 
depends on the initial state, like a small positive 
initial velocity or force.

%----------------------------------------------------------
\begin{figure}[!t]
\centering
\includegraphics[width=0.75\textwidth]{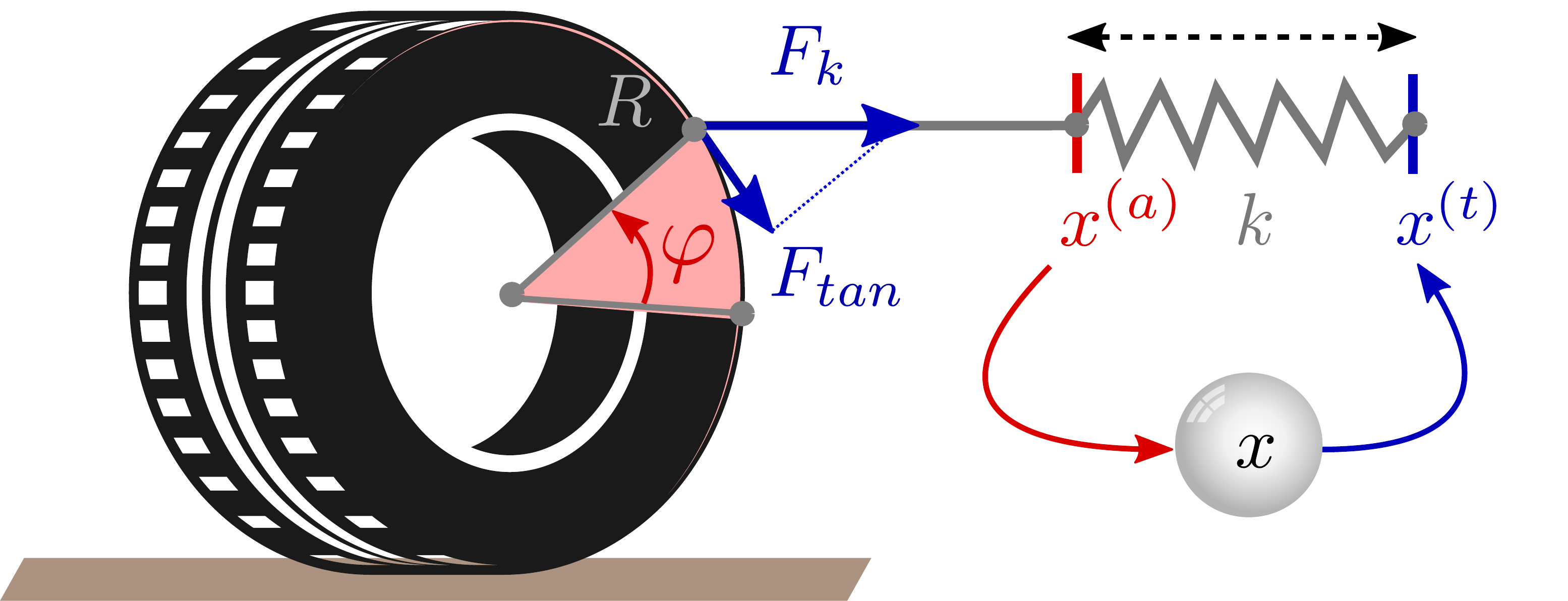}
\caption{Illustration of a one-neuron controller simulating 
the transmission of classical steam engines. The actual
position $x^{(a)}=\cos(\varphi)$ of the wheel drives,
as described by (\ref{dot_x}), the neural
activity $y(x)$ setting the target position $x^{(t)}=y$. 
A simulated spring with spring constant $k$ between 
$x^{(a)}$ and $x^{(t)}$ generates subsequently the
torque $RF_\mathrm{tan}$ acting on the wheel. Here 
$F_\mathrm{tan}=F_k\sin(\varphi)$ denotes the tangential 
projection of the spring force $F_k=k(x^{(t)}-x^{(a)})$.
}
\label{fig:steamEngineController}
\end{figure}
%----------------------------------------------------------

% ------------
\subsection{Locomotion via time reversal symmetry breaking}
% ------------

Locomotion is parametrized typically by a velocity vector
$\mathbf{v}=d\mathbf{r}/dt$ that incorporates both 
the direction and the magnitude of the movement.
Reversing time $t\leftrightarrow (-t)$ reverses then
also the velocity vector. Here we are interested in 
self-organized robots that break time reversal symmetry 
spontaneously, which in our case implies  that the 
attracting states in the sensorimotor loop come in pairs
that are related via time reversal symmetry. Whether
the robot moves for- or backwards depends then only
on the initial conditions. For this purpose we
use the one-neuron controller illustrated
in Fig.~\ref{fig:steamEngineController}.

A wheel with a rotational angle $\varphi$ 
is regulated individually via
\begin{equation}
\tau \dot x = \cos\varphi -x, 
\qquad\quad y=\tanh(ax)~,
\label{dot_x}
\end{equation}
where $x$ is the membrane potential of the controlling
neuron, $y=\tanh(ax)$ the neural activity and $\tau$ 
the membrane time constant. The motor command is 
proportional to the spring force
\begin{equation}
F_k=k\big(x^{(t)}-x^{(a)}\big),
\qquad\quad
x^{(t)}= y,
\qquad\quad
x^{(a)}=\cos(\varphi)\,,
\label{F_k}
\end{equation}
where $k$ is a spring constant and $x^{(a)}$ and
$x^{(t)}$ respectively the actual and the target 
position of the wheel in terms of a projection to 
the ground \cite{sandor2018kick}. Note that the 
angle $\varphi$, which enters the right-hand side 
of Eqs.~(\ref{dot_x}) and (\ref{F_k}) as $\cos(\varphi)$, 
is the measured, the actual angle of the wheel. All 
forces, gravitational and mechanical, impact the 
controller hence exclusively via their influence on 
the angle $\varphi$.

The controller simulates the transmission rod of 
a classical steam engine, as sketched in 
Fig.~\ref{fig:steamEngineController}, as it
translates the bounded forth and back motion 
of the neural activity $y(t)$ into a rotational 
motion. Alternatively, instead of using the
angle $\varphi$ as the determining variable, one 
could postulate a discrete map $\omega^{(t)}=\tanh(a\omega^{(a)})$
between the actual and a target angular velocity
\cite{der2008predictive}, $\omega^{(a)}$ and 
$\omega^{(a)}$. This is not a problem for simulated
robots, for which $\omega$ is a directly accessible 
variable. To obtain a reliable estimate of the instantaneous
angular velocity for real-world robots working with 
duty cycles of the order of $20\,\mathrm{Hz}$
would however be a challenge \cite{sandor2018kick}. 

A controller enabling locomotive limit cycles to
emerge in the sensorimotor loop \cite{martin2016closed},
as described here by (\ref{dot_x}) and (\ref{F_k}), 
differs qualitatively from controlling schemes employing
local phase oscillators \cite{ambe2018simple}, for which 
a spontaneous reversal of the direction of motion would 
not be possible.

% ------------
\subsection{Isolated wheel}
% ------------

The individual wheels of the simulated robots are 
controlled exclusively by (\ref{dot_x}) and 
(\ref{F_k}). There is no explicit inter-wheel
coupling present. It is illustrative to model,
for comparison, an idealized isolated wheel
with moment of inertia $I$, radius $R$, angle $\varphi$ 
and angular velocity $\omega$. The force $F_k$
generated by the simulated transmission rod then
enters the equations of motion as a torque 
$RF_k\sin(\varphi)$,
\begin{equation}
\tau \dot x = \cos\varphi -x,
\qquad\quad
\dot\varphi= \omega,
\qquad\quad
I\dot\omega = R(F_k\sin\varphi -f \omega)~,
\label{eq_dot_x_phi_omega}
\end{equation}
where $f>0$ is a friction coefficient. 
Eq.~(\ref{eq_dot_x_phi_omega}) is manifestly 
invariant under $\omega\leftrightarrow(-\omega)$,
$\varphi\leftrightarrow(-\varphi)$ and
$x\leftrightarrow x$, which implies
time-reversal symmetry in terms of an
invariance with respect to reversing 
the direction of motion. We will investigate
(\ref{eq_dot_x_phi_omega}) further in
Sect.~\ref{sec_theory}, noting here that
symmetry breaking may occur also in embodied 
robots that incorporate forward world 
models \cite{der2013behavior}.

%----------------------------------------------------------
\begin{figure}[!t]
\centering
\includegraphics[height=0.4\textwidth]{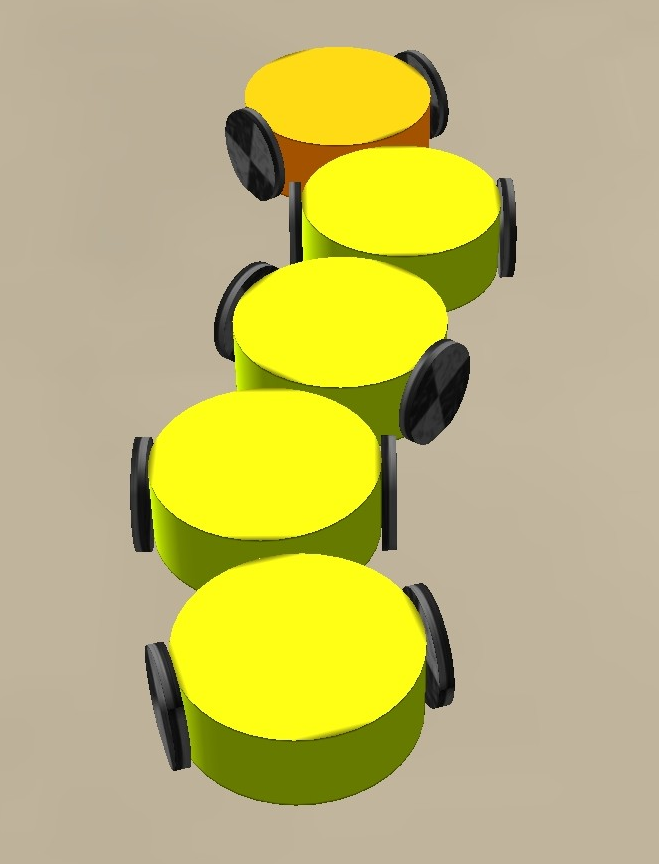}
\hspace{4ex}
\includegraphics[height=0.4\textwidth]{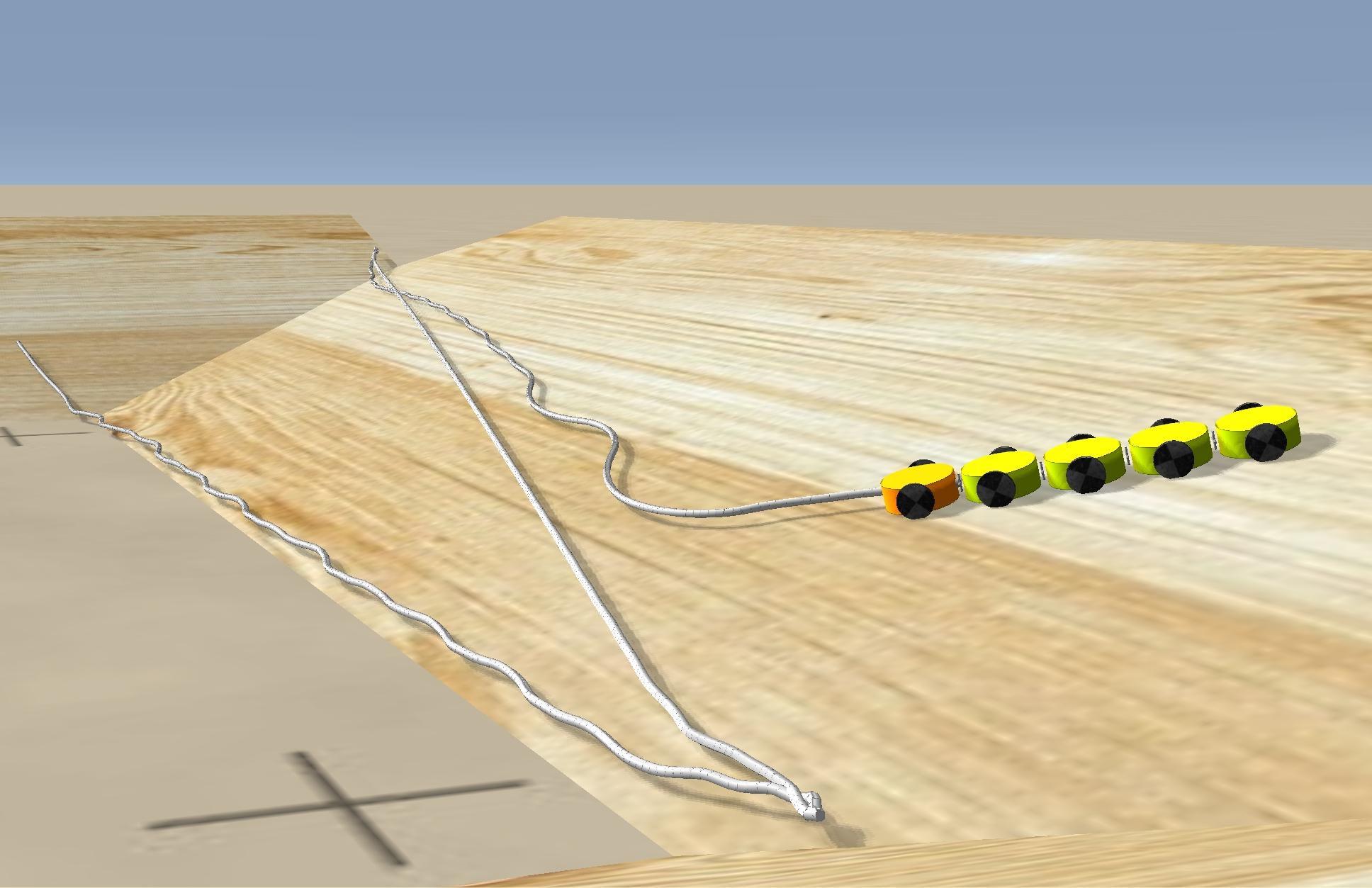}
\caption{Screenshots of the LPZRobots simulation
environment. {\it Left:} A snake-like train of cars composed 
of five passively coupled segments. Each segment contains
two independent wheels that influence each other exclusively 
through the mechanics of the body and via the hinge joints
connecting the individual segments.
{\it Right:} A robot climbing intersecting slopes on its 
own, with the silver line illustrating the ground 
trace of the last segment. No explicit control signal has 
been given. The wiggling observed when the robot moves fast 
on straight stretches disappears at lower velocities, as it 
is the case when moving steeper up the slope. The train of 
cars reverses direction autonomously when hitting the intersecting
slope. The wiggling amplitude on the last leg increases progressively 
while moving down, leading in the end to an upward curve
(\href{http://doi.org/10.6084/m9.figshare.7643123.v1}
{click for movie}).
}
\label{fig:screenshots}
\end{figure}
%----------------------------------------------------------
%%%%%%%%%%%%%%%%%%%%%%%%%%%%%%%%%%%
\section{Results}
%%%%%%%%%%%%%%%%%%%%%%%%%%%%%%%%%%%

We used the LPZRobots physics simulation
package \cite{der2012playful} for the simulation
of robots composed of chains of 1-5 two-wheeled
cars linked passively through hinge joints, which
are equipped with passively damped torsion springs. 
In the absence of motor commands or external
forces the equilibrium position of the hinge joints
induces a straight alignment of the connected body 
segments. During locomotion the joints can store,
on the other hand, potential energy when bent.

Shown in Fig.~\ref{fig:screenshots} is the trajectory 
of the train of cars climbing up a slope that is 
intersected orthogonally by two other slopes. One 
observes wiggling and straight locomotion together 
with direction reversal and large turns.
In order to develop an understanding we start 
by investigating the velocity profile of a robot on an 
extended slope, concentrating on the dependence of the 
self-regulated steady state velocity on the spring 
constant $k$ of the actuator and on the inclination of 
the slope. We note that the simulation cycle times of 
the LPZRobots simulation package, which is based
on the Open Dynamics Engine \cite{smith2005open},
are of the order of 50\,ms. 

% ------------
\subsection{Moving up and down an infinite slope}
% ------------

\begin{figure}[!t]
\centering
\includegraphics[width=0.80\textwidth]{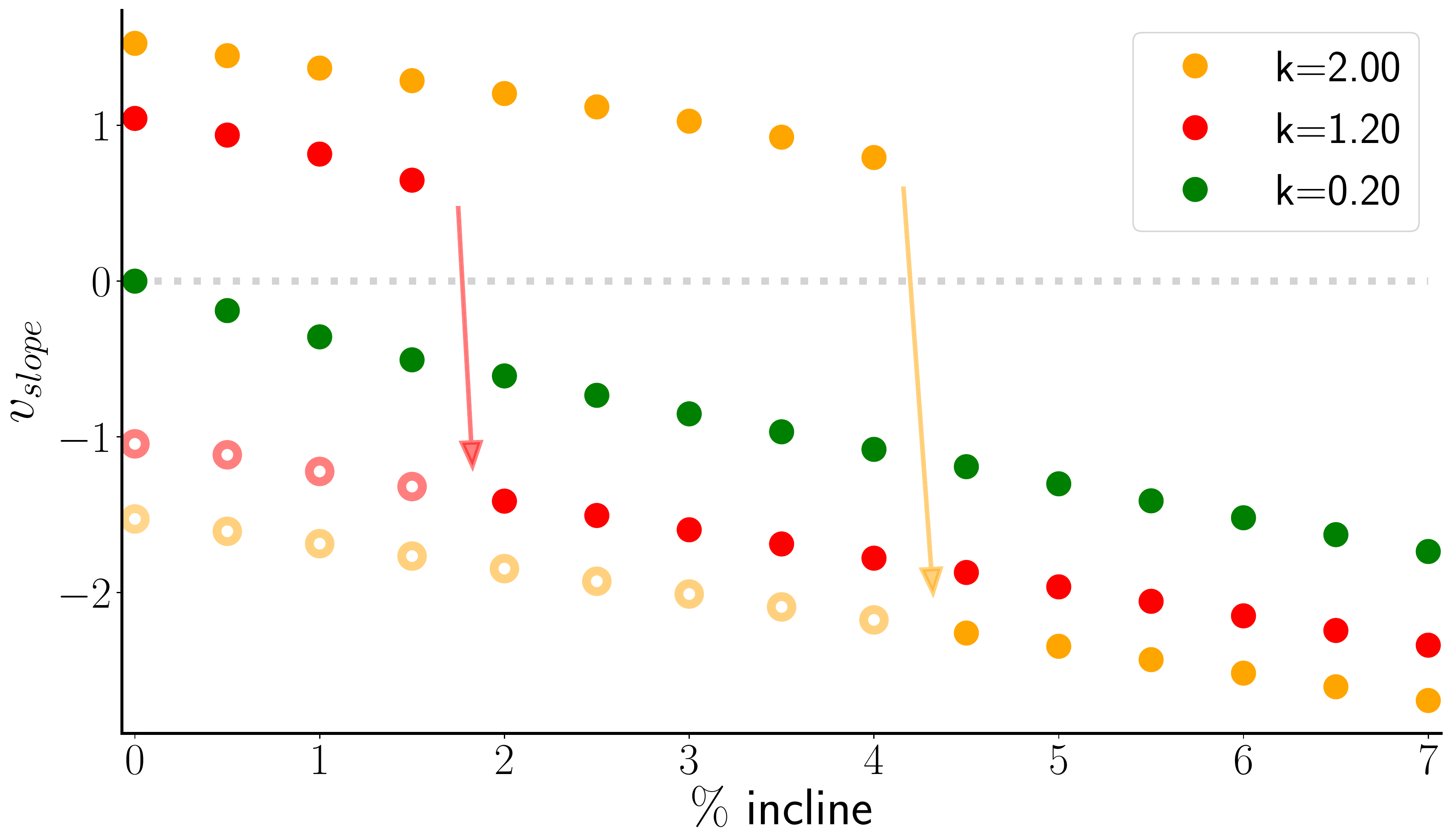}
\caption{The velocity profile of a 5-segmented robot on
a slope. The parameters are $a=1$ and 
$k=2.0/1.2/0.2$ (yellow/red/green). Shown are the 
steady-state velocities for moving directly upwards 
(positive $v_{slope}$) and for moving down the slope 
(negative $v_{slope}$). A first order transition occurs
for $k=2.0$ (yellow dots) and $k=1.2$ (red dots) when 
increasing the inclination slowly, that is adiabatically, 
as indicated by the arrows. The robot cannot move upwards 
for $k=0.2$ (green dots), being subcritical.
}
\label{fig:velocities}
\end{figure}

In Fig.~\ref{fig:velocities} we present the velocity
profile for a 5-segmented robot moving on a slope
parallel to the gradient, that is straight up and down.
The downward velocity decreases in magnitude with 
decreasing slope and spring constant $k$, as expected.

For the robot moving on a horizontal plane there
exists a critical $k_c\approx0.54$, such that
the limit-cycles corresponding to regular forward 
or backward movement disappear for $k<k_c$. For
a spring constant of $k=0.2$, which is below $k_c$,
the robot moves therefore only when the slope has a
finite downward inclination, as shown in 
Fig.~\ref{fig:velocities}.

%----------------------------------------------------------
%----------------------------------------------------------

The torque exerted on the wheels is directly proportional 
to the spring constant, being generated otherwise through
the sensorimotor feedback. Moving upward the slope the
torque $R F_{tan}$, and hence also the sensorimotor feedback, 
needs to counter the gravitational downhill force $F_G$. For 
an engine producing a constant torque, the balancing of the
tangential and the gravitational force would lead to an uphill 
velocity $v_{slope}\propto F_{tan}-F_G$ that vanishes
linearly and hence continuously at a critical inclination.
This is however not the case when the motion is generated 
through sensorimotor feedback, as evident from the data 
presented in Fig.~\ref{fig:velocities}. The sensorimotor
feedback involves a self-consistency condition that
breaks down discontinuously, at finite values of the 
uphill velocity, when the inclination of the slope becomes 
too large. It remains however the case that larger
spring constants $k$ allow the robot to move up steeper
slopes.
  
% ------------
\subsection{Autonomous direction reversal}
% ------------

The trajectory of the robot presented in
Fig.~\ref{fig:screenshots} hits twice an 
intersecting slope. A robot equipped with
actuators producing constant torques would
move segment by segment onto the intersecting 
slope, up to the point where the gravitational 
downward pull of the increasing number of segments 
cancels with the locomotive force. At this point
the robot would remain in place.

The locomotive force of the snake-like robot presented 
here is however highly compliant, being generated 
within the sensorimotor loop. The velocity profile
presented in Fig.~\ref{fig:velocities} implies 
that an equilibrium position resulting from the
balance of an upward locomotive force and the
downhill gravitational pull is not possible, namely 
that the limit-cycle attractors present in the 
sensorimotor loop allow the robot to move only 
up- or downhill. We note that the stable fixpoint
attractors corresponding to a non-moving state,
that exist in conjunction with the limit-cycle 
locomotion for an isolated wheel, as discussed in 
Sect.~\ref{sec_theory}, possesses only a vanishing
small basin of attraction for the case of a train
of cars. A robot hitting a slope that is too steep 
is therefore likely to reverse direction, as observed in 
Fig.~\ref{fig:screenshots}, instead of being pulled
into a fixpoint attractor and coming to a stop.

%----------------------------------------------------------
\begin{figure}[!t]
\centering
\includegraphics[height=0.75\textwidth]{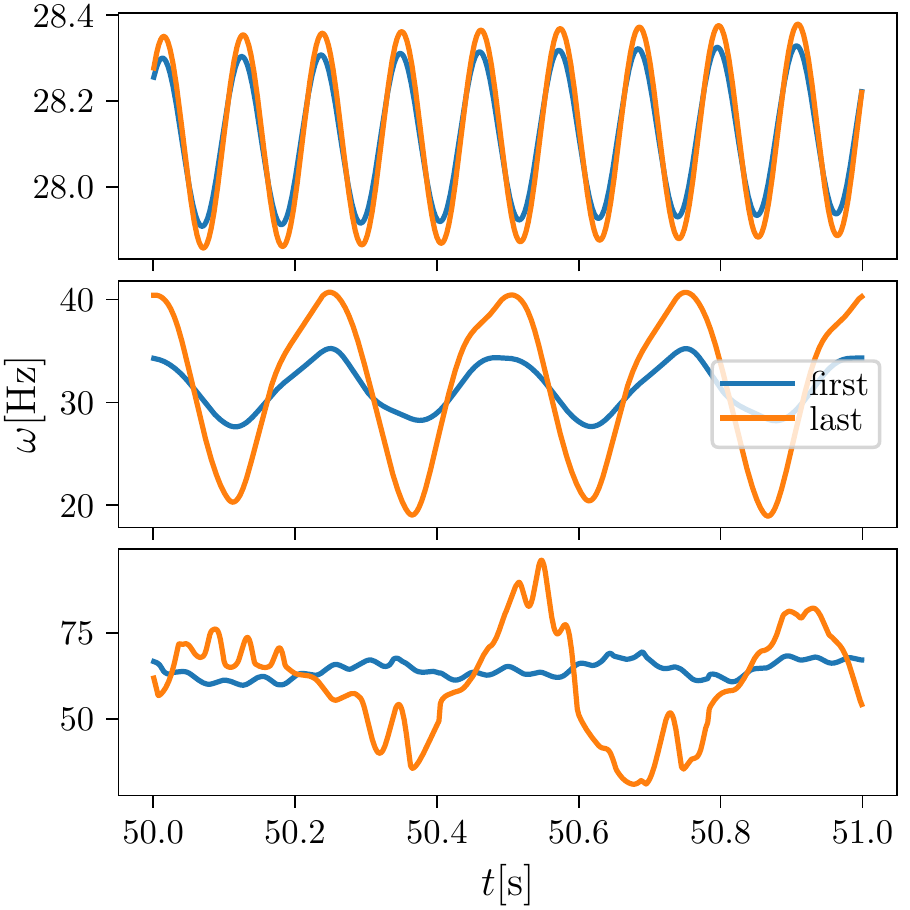}
\caption{As a function of time, the measured angular velocity 
$\omega$ of one of the two wheels of the first and of the last
car. The membrane time constant is $\tau=50\,\mbox{ms}$. 
{\it Top:} For $a=0.5$ and $k=0.5$, a straight-moving mode
%(\href{https://itp.uni-frankfurt.de/~gros/Movies/wheelRobot/carChain_straight_tilting.mp4}
Left and right wheels of a given car are exactly synchronized.
The small difference in amplitude in the $\omega(t)$ oscillations
between cars is due to small oscillations in the respective
pitch angles. Note that $\omega(t)$ is not exactly constant
as resulting from (\ref{dot_x}).
{\it Mid:} For $a=1.0$ and $k=1.0$, a regular meandering mode 
%(\href{https://itp.uni-frankfurt.de/~gros/Movies/wheelRobot/carChain_meandering_mode.mp4}
%{click for movie}).
{\it Bottom:} For $a=1.0$ and $k=5.0$, a chaotic mode resulting from wide
sideway swings of the tail
%(\href{https://itp.uni-frankfurt.de/~gros/Movies/wheelRobot/carChain_chaotic_mode.mp4}
%{click for movie}).
(\href{https://doi.org/10.6084/m9.figshare.8143040.v1}{click for movies}).
}
\label{fig:time_series}
\end{figure}
%----------------------------------------------------------

%----------------------------------------------------------
\begin{figure}[!t]
\centering
\includegraphics[height=0.27\textwidth]{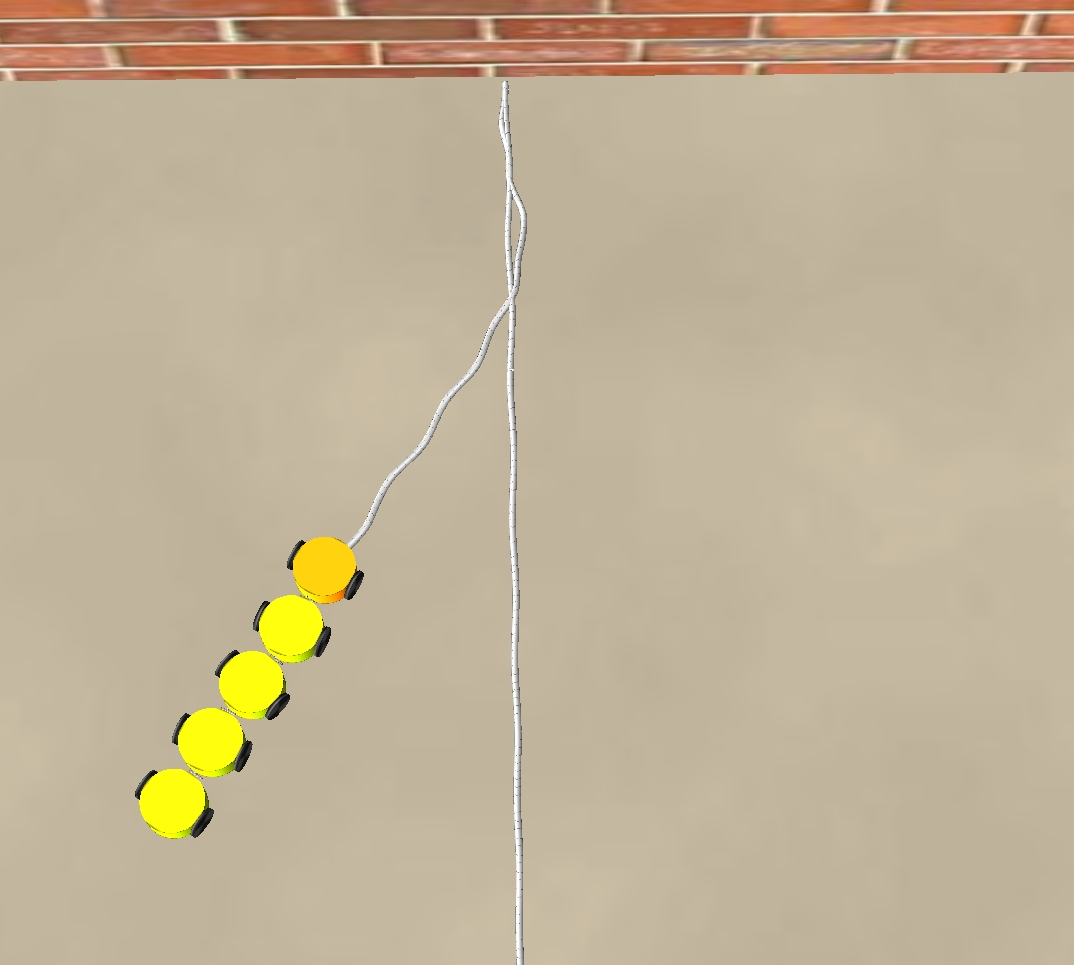}
\hfill
\includegraphics[height=0.27\textwidth]{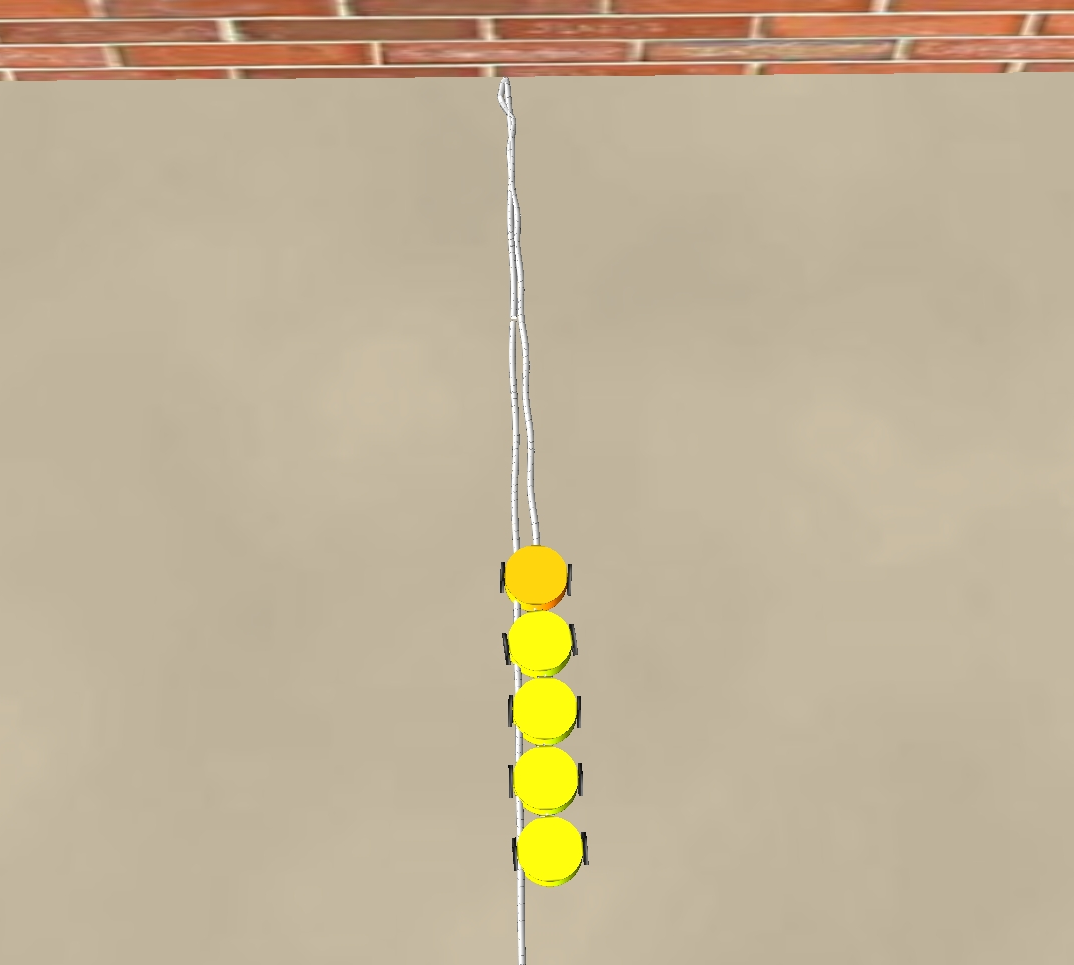}
\hfill
\includegraphics[height=0.27\textwidth]{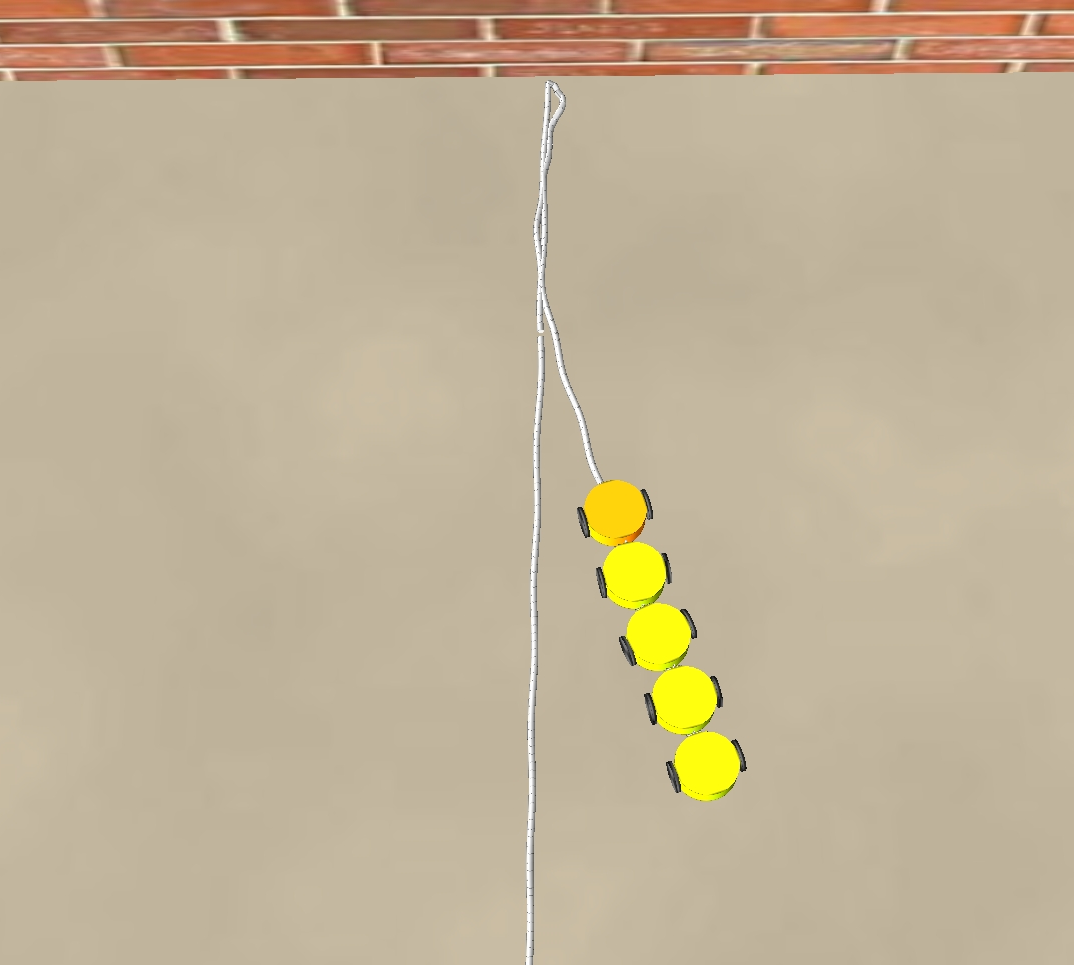}
\caption{A five-segmented robot bouncing off the wall, with
the silver line tracing the position of the darker segment.
The only knowledge the robot disposes of the outside world,
and hence of the presence of a wall, is via propriosensation, 
that is via the measurement of the angle of the wheels. The
slight wiggling of the forward motion causes the robot to be 
reflected at various angles, even though it bumps into the wall 
perpendicularly (with respect to the average direction
of locomotion).
The direction reversal results from the destruction of the
forward limit cycle upon hitting the wall, with the flow
in phase space evolving subsequently towards the backward
attractor.
}
\label{fig:train_wall_bounce}
\end{figure}
%----------------------------------------------------------

% ------------
\subsection{Straight, meandering and chaotic modes}
% ------------

The last leg of the trajectory shown in 
Fig.~\ref{fig:screenshots} shows growing
left- and rightward swings. This is a typical
behavior at larger velocities, here due to 
the slight downhill direction, that results 
from a transversal mechanical instability of 
the connected segments. The joints are elastic
and therefore capable to store a certain amount
of energy, akin to what happens when a string starts
to vibrate. The final upturn of the robot
may occur at an angle, interestingly, that 
puts the robot below criticality. The robot will
then stop moving and reverses direction.

%---------------------------------------------------------
\begin{figure}[!t]
\centering
\includegraphics[width=0.75\textwidth]{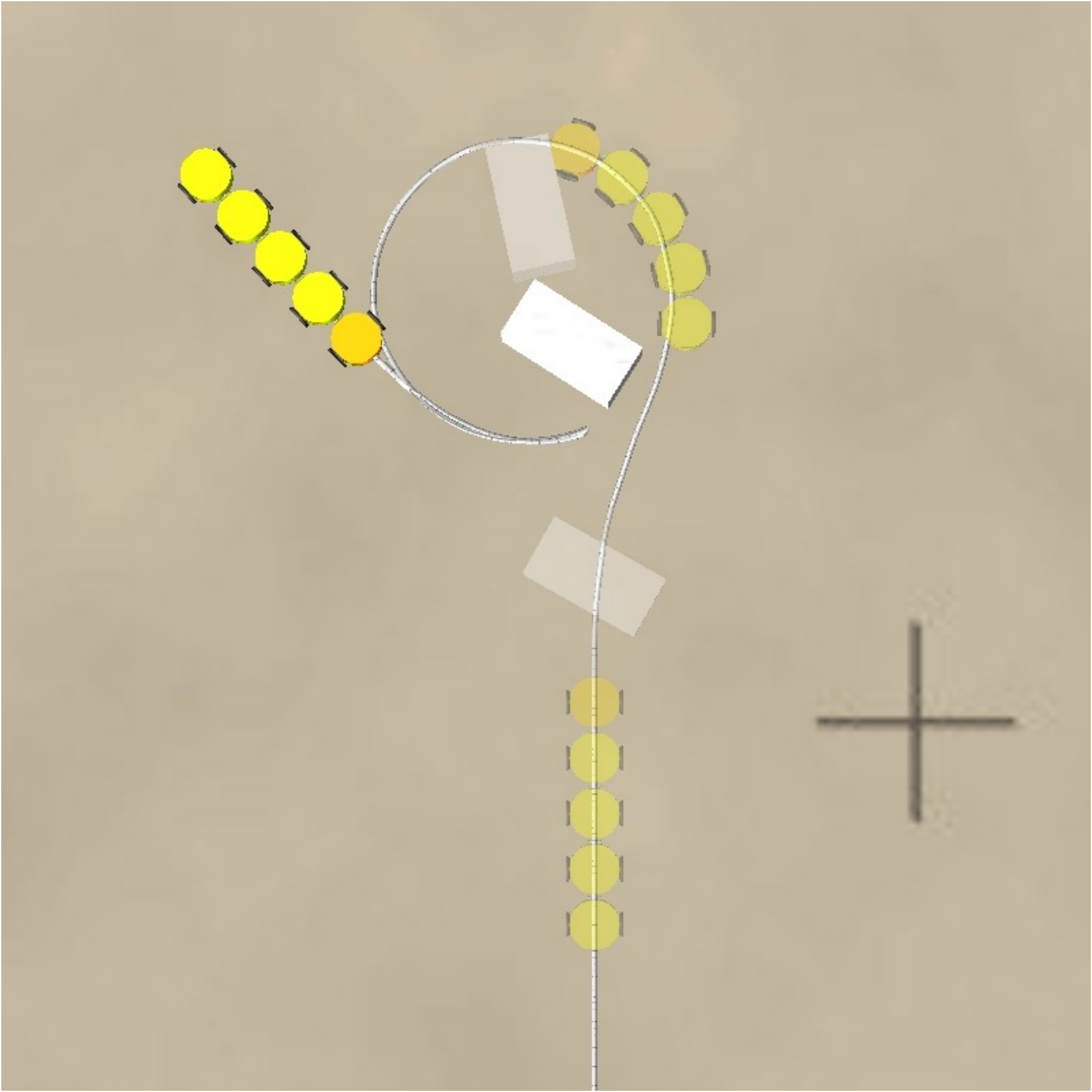}
\caption{Superimposed screenshots from the LPZRobots simulation
environment, with $a=1$ and $k=2$. The silver line represents 
the ground trace of the darker of the two end segments. The 
snake-like train of cars
starts from the lower center (shadowed), where it first hits 
a movable box (shadowed), bending and pushing the box around in an 
inward spiral (as indicated by the shadowed box and robot in 
the middle). As the angle of the spiral becomes steeper and the 
forward velocity smaller, the robot will reach the point at which
the forward limit cycle disappears, as seen also in
Fig.~\ref{fig:velocities}. The robot then stops for a short
period during which the dynamics flows in phase space 
autonomously towards the attractor corresponding to backward 
motion. Once reached, the robot reverses direction, as one 
can see from the positioning of the dark end segment
(\href{https://doi.org/10.6084/m9.figshare.8143040.v1}{click for movie}).
}
\label{fig:train_boxes}
\end{figure}
%----------------------------------------------------------

In Fig.~\ref{fig:time_series} we present the angular 
frequency $\omega(t)$ for one of the two wheels of the 
first and the last car, respectively, of a five-segmented
snake-like robot moving on a flat ground. Shown are the
timelines for two regular and for a chaotic mode. In the 
first limit-cycle mode the train of cars moves straight. 
The ten wheels are in this case synchronized in the sense
that the small modulation of the respective angular 
velocities, which occur because the controller (\ref{dot_x}) 
is not rotationally invariant, appear all at exactly the 
same time. Their respective amplitudes vary however from
car to car, which implies that the pitch angles of the 
individual cars oscillates, even though only slightly.

For larger average velocities, the limit-cycle straight 
mode tends to be become unstable, making way to
meandering and chaotic modes, as illustrated in
Fig.~\ref{fig:time_series}. We did not investigate
the exact nature of the respective transition to chaos, 
which may be due to a cascade of period-doubling within 
the space of meandering modes \cite{gros2015complex}.
We also note that our classification of the highly
irregular mode as chaotic relies here only on a visual 
inspection, as we did not apply a formal test for the 
presence of deterministic chaos \cite{wernecke2017test}.

% ------------
\subsection{Interacting with a structured environment}
% ------------

The robots exhibit interesting behavioral patterns when 
situated in a structured environment. As a first example 
we show in Fig.~\ref{fig:train_wall_bounce} the interaction
of a 5-segmented robot with a wall. Before hitting the
wall the robot possesses, due to time-reversal symmetry,
both a forward- and a backward-moving limit cycle.
Approaching the wall the sensorimotor state of the robot 
is in the forward limit cycle, which becomes however 
destroyed upon hitting the wall. The flow in phase space,
that is the evolution of the membrane potentials
$x$ of the individual wheels, is then attracted by
the remaining limit cycle, which is the one corresponding 
to moving backward. The robot hence reverses direction.

The observed direction reversal occurs autonomously
in the absence of top-down control signals. As the robot possesses
only sensors measuring the angles of the wheels there 
are furthermore no external sensors present that would
inform the robot about the distance to the obstacle. One 
observes, as shown in Fig.~\ref{fig:train_wall_bounce}, that 
the angle at which the robot bounces varies considerably as a
consequence of the wiggling of the initial forward motion.
In a slow-velocity non-wiggling mode the bouncing occurs
exactly perpendicularly.

We also studied the interaction of the multi-segmented robot with
movable boxes, as illustrated in Fig.~\ref{fig:train_boxes}.
Due to the mechanical feedback of the passive but elastic
hinge joints, the body of the train of cars bends, such that 
the robot continues to push the box in ever smaller circles.
During the push, the robot slows down continuously, until
a critical velocity is reached and the forward limit cycle
disappears. At this point the robot stops,
reversing direction autonomously.

The robot disposes of a single controlling neuron per actuator, 
which obtains in turn information about the external world only
indirectly, namely via the measured angle of the respective
wheel. The behavior illustrated in Fig.~\ref{fig:train_boxes},
such as pushing around boxes, is hence emergent and an example 
that embodied controlling frameworks may lead robustly to novel 
behavioral patterns. Functionally, the ability of the snake-like
robot to follow spiral-shaped trajectories, when pushing a box, 
is a direct consequence of the highly elastic working regime of 
the actuators, viz of compliance. By themselves, the wheels on 
the left-hand and the right-hand side of the body would acquire, 
being subject to identical controllers, identical angular velocities. 
The fact, however, that the turning speed of the wheels is not determined 
by a top-down signal, but by the sensorimotor feedback, allows the 
emerging limit cycles to adapt autonomously to environmental
forces. The wheels on opposite sides of the body react
as a consequence distinctly when the box exerts a non-symmetric 
resistance onto the robot. The respective angular velocities then 
differ.

% ------------
\subsection{Theory for an isolated single wheel}\label{sec_theory}
% ------------

The fixpoints of an isolated and non-moving single wheel, 
as described by Eq.~(\ref{eq_dot_x_phi_omega}), are determined 
by $\omega=0$, $x=\cos \varphi$ and
\begin{equation}
\sin\varphi=0 \qquad\mbox{or}\qquad y(x)=\cos\varphi\,,
\label{omegaNull}
\end{equation}
compare Eqs.~(\ref{eq_dot_x_phi_omega}) and 
(\ref{F_k}). The Jacobian $J(\varphi)$
is in general
\begin{equation}
J(\varphi)\ =\ 
\left(\begin{array}{ccc}
-1/\tau & \,-\sin\varphi/\tau\, &0 \\[0.5ex]
0 & 0 & 1 \\[0.5ex]
ka(1-y^2)\sin\varphi & A & -f
\end{array}\right)\,,
\label{Jacobian}
\end{equation}
where $A=ky\cos\varphi+k(\sin^2\varphi-\cos^2\varphi)$.
We have set $R=1$, measuring in addition $k$ and
$f$ relative to $I$.

\begin{itemize}
\item[--] For the first fixpoint $\sin\varphi=0$, that
          is for $(x,\omega,\varphi)=(1,0,0)$ and
          $(-1,0,\pi)$, one has 
          $A(0,\pi)=k(\pm y(\pm 1)-1)=k(\tanh(a)-1)$, 
          which is always negative. The eigenvalues of 
          the Jacobian are then
\begin{equation}
\lambda_0(0,\pi)=-\frac{1}{\tau},\qquad\quad
\lambda_\pm(0,\pi)=-\frac{f}{2}\pm\frac{1}{2}\sqrt{f^2+A(0,\pi)}\,,
\label{eq_fixpoint_0_pi}
\end{equation}
Since $A(0,\pi)<0$, the fixpoints at $\varphi=0$ and 
$\varphi=\pi$ are always stable.

\item For the second set of fixpoints one has to solve the
self-consistency condition $y(x)=\tanh(ax)=x$,
with $x=\cos\varphi$. The trivial solution
$x=0$, that is $(x,\omega,\varphi)=(0,0,\pm\pi/2)$ 
splits at $a_\mathrm{c}=1$ via a pitchfork transition,
allowing for three coexisting fixpoints for $a>a_c$.
For small $k$ the trivial fixpoint $x=0=\cos\varphi=y$ 
with the Jacobian
\begin{equation}
J(\varphi=\pm\pi/2)\ =\ 
\left(\begin{array}{ccc}
-1/\tau & \,\mp 1/\tau\, &0 \\[0.5ex]
0 & 0 & 1 \\[0.5ex]
\pm ka & k & -f
\end{array}\right)\,
\label{Jacobian_x_0}
\end{equation}
is  a saddle for $0<a\le1$, having two 
negative and one positive eigenvalues, being
stable however for $a>1$. For larger values of
the spring constant~$k$, the $x=0$ solution undergoes 
a Hopf bifurcation leading to limit-cycle oscillations.
\end{itemize}

%---------------------------------------------------------
\begin{figure}[!t]
\centering
\includegraphics[width=0.95\textwidth]{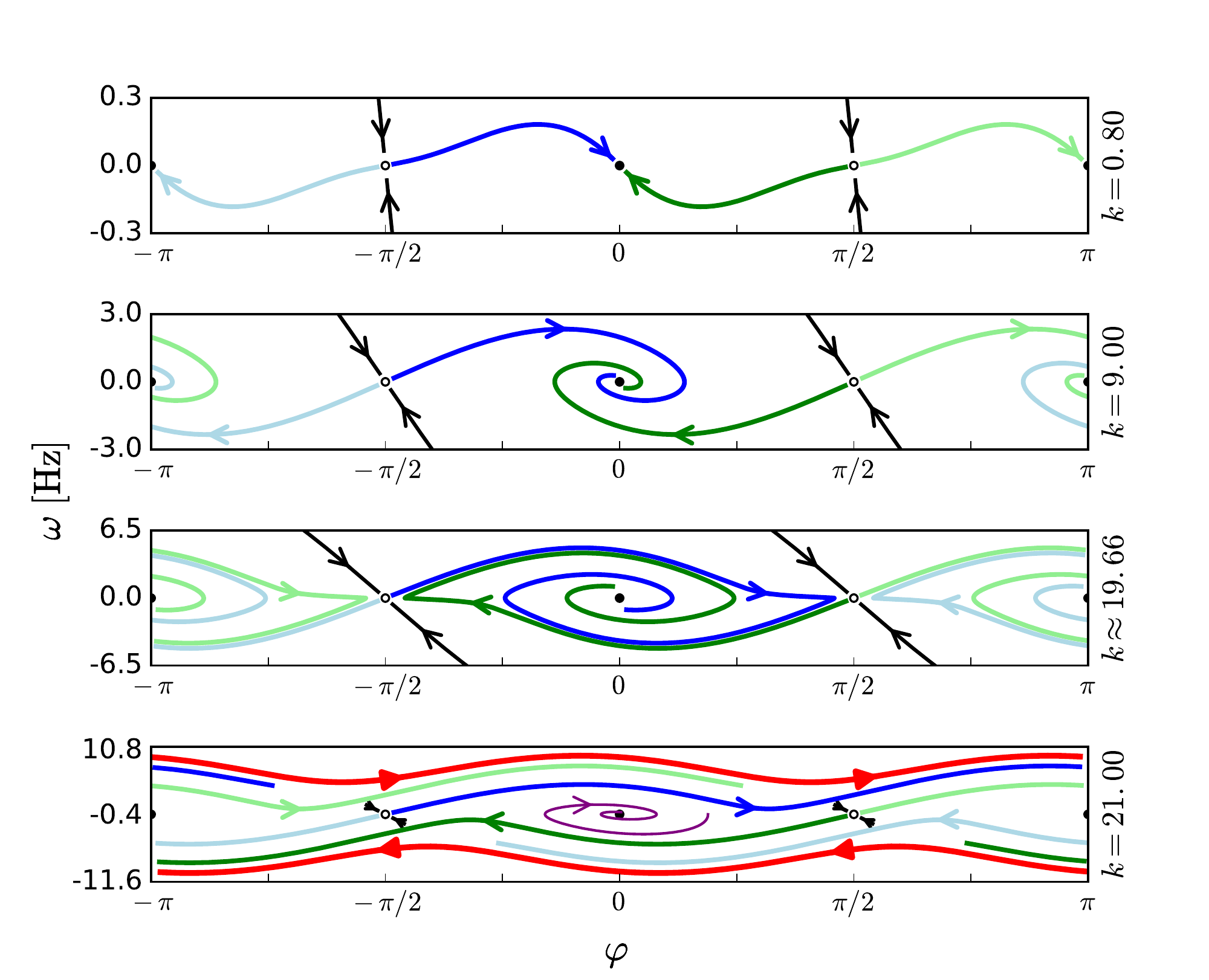}
\caption{One-step heteroclinic route to locomotion for $a<1$.
Shown are stable limit cycles (red) and stable, unstable 
manifolds   (black, light/dark blue/green) and 
selected sample trajectories (violet). The fixpoints at 
$\varphi=0,\,\pi$ are stable nodes/foci respectively for 
small and larger spring constants $k$ (top and upper middle panel), 
with the saddles at $\varphi=\pm\pi/2$ remaining 
unchanged in character for all $k$. A symmetric heteroclinic 
connection between the saddles is created when increasing 
$k$ further. For $k\approx19.66$ (lower bottom panel)
a stable limit cycle (red) corresponding to limit-cycle 
locomotion (bottom panel) is generated. 
The parameters are $a=0.95$, $\tau=0.2$, $I=0.25$ and $f=0.5$.
}
\label{fig:bifurcations_a_smaller_1}
\end{figure}
%----------------------------------------------------------

%---------------------------------------------------------
\begin{figure}[!t]
\centering
\includegraphics[width=0.95\textwidth]{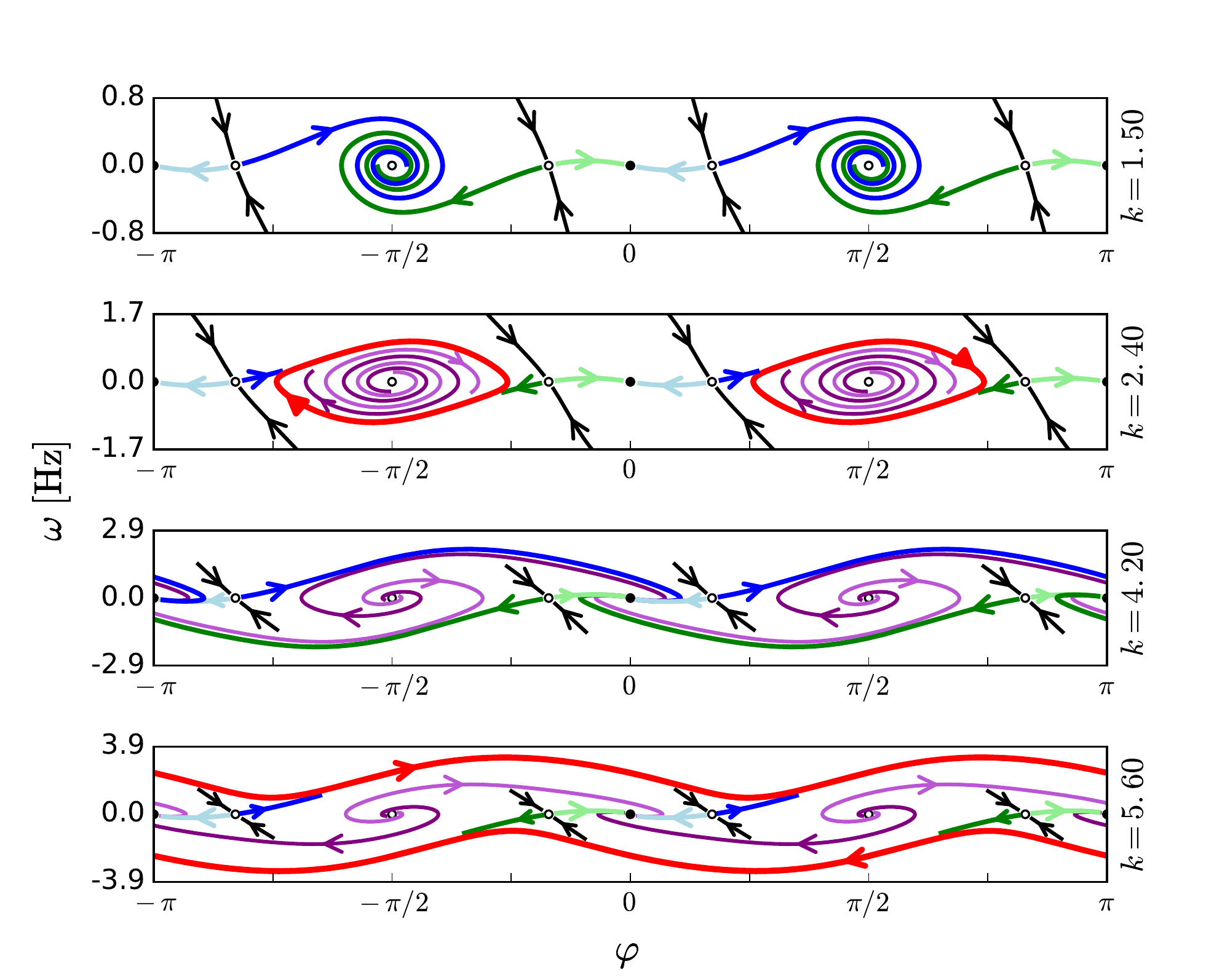}
\caption{Multi-step heteroclinic route to locomotion for $a>1$.
The fixpoints at $\varphi=0,\pi$ are always stable nodes, 
with the foci at $\varphi=\pm\pi/2$ undergoing a Hopf
bifurcation upon increasing the spring constant $k$
(top/upper middle panel). The resulting limit
cycle (red) corresponds to a periodic forth-and-back
motion characterized by $\omega\ne0$ and a vanishing
average $\overline{\omega}=0$. The forth-and-back
motion is destroyed by a symmetric heteroclinic transition
(upper/lower middle panel), leading to an intermediate
phase without locomotion. A second heteroclinic transition
then generates stable limit-cycle locomotion 
(lower middle/bottom panel). Parameters, besides $a=1.5$, 
and color-coding as for Fig.~\ref{fig:bifurcations_a_smaller_1}.
}
\label{fig:bifurcations_a_larger_1}
\end{figure}
%----------------------------------------------------------

% ------------
\subsection{Routes to locomotion}
% ------------

It is interesting to study how limit-cycle
locomotion arises from a configuration of
individual fixpoints upon increasing the
force acting on the wheel, that is the spring 
constant $k$.

In Fig.~\ref{fig:bifurcations_a_smaller_1} we illustrate
the case $a<1$, for which the saddles at 
$\varphi=\pm\pi/2$ do not undergo a pitchfork bifurcation 
yet.  Locomotion arises in this case via a one-step 
heteroclinic transition which allows for the generation 
of limit cycles of finite amplitudes. A pair of stable
and unstable limit cycles is eventually produced
when increasing the spring constant $k$, with the 
stable limit cycle corresponding to a locomotive 
behavioral primitive. Note that the phase space
of (\ref{eq_dot_x_phi_omega}) is three dimensional 
and that the flow shown in 
Fig.~\ref{fig:bifurcations_a_smaller_1} corresponds
to a projection onto the $(\varphi,\omega)$-plane.
Trajectories may hence intersect.

In Fig.~\ref{fig:bifurcations_a_larger_1} 
a multi-step route to locomotion for $a>1$ is presented. 
One observes first a Hopf-bifurcation (HB) at
$\varphi=\pm\pi/2$, which leads to a first 
intermediate phase characterized by a closed 
limit-cycle in the $(\varphi,\omega)$-plane. 
This limit cycle, corresponding to 
small amplitude forth-and-back periodic motion, 
is destroyed when hitting in a symmetric heteroclinic 
transition (SHE) the two saddles present additionally
for $a>a_\mathrm{c}=1$. Motion ceases in the subsequent 
second intermediate phase, for which the 
unstable trajectories emerging from 
$\varphi=\pm\pi/2$ lead to the fixpoints~$\varphi=0,\pi$. 
A stable limit-cycle emerges however from a second 
heteroclinic transition (HE) when increasing the spring
constant $k$ further, namely when the unstable
manifold of one of the additional saddles
hits another saddle.

For the parameters used for Fig.~\ref{fig:bifurcations_a_larger_1} 
there are hence for $a>1$ two phases without locomotion,
viz for which the angular frequency $\omega$ decays to 
zero. The average angular frequency $\overline{\omega}$ 
vanishing for forth-and-back motion, but not for limit-cycle
locomotion:
$$
\fbox{$\phantom{|}\omega\to0\phantom{|}$} 
\ \ \xrightarrow{\mbox{\small HB}} \ \
\fbox{$\phantom{|}\omega\ne0,\,\overline{\omega}=0\phantom{|}$} 
\ \ \xrightarrow{\mbox{\small SHE}} \ \
\fbox{$\phantom{|}\omega\to0\phantom{|}$} 
\ \ \xrightarrow{\mbox{\small HE}} \ \
\fbox{$\phantom{|}\omega\ne0,\,\overline{\omega}\ne0\phantom{|}$} 
$$
With the motor command being proportional to the spring
constant $k$, it is somewhat intuitive that one needs a 
critical $k$ for locomotion to emerge. Relatively large 
spring constants have been used in 
Fig.~\ref{fig:bifurcations_a_smaller_1} for 
illustrative purposes.

%%%%%%%%%%%%%%%%%%%%%%%%%%%%%%%%%%%
\section{Conclusions}
%%%%%%%%%%%%%%%%%%%%%%%%%%%%%%%%%%%

The gait of an animal corresponds to a coordinated
pattern of limb movements that repeats with a certain 
frequency. Gaits are generated typically by a central 
pattern generator \cite{hopkins2009survey},
that is by a central processing unit that produces
coordinated motor signals. We have examined here an 
alternative framework for which the actuators of an 
animat are self active, with the dynamics of the 
individual actuators resulting from the presence 
of limit-cycle attractors within the sensorimotor loop.
The actuators, in our case the wheels of a snake-like
robot, are coupled in this framework only via the
mechanics of the body and by the reaction of the environment.
The gaits of the animat result in our framework therefore 
from self-organizing principles. For a simulated 
wheeled snake-like animat, we find that the robot
interacts autonomously with the environment, e.g.
by turning on its own on a slope. The robot will also
push a movable box around for a while when colliding 
with one.

We have tested in addition that locomotion modes also 
arise for heterogeneous (not identical) body segments, 
e.g.\ when the controllers have different spring constants 
$k_i$, or different wheel sizes. The multi-segmented robot 
is capable of generating locomotion in particular when
several actuators are subcritical, i.e.\ with wheels 
which on their own would not maintain oscillatory 
dynamics. Embodiment leads in our study robustly to 
emergent locomotion.

The here employed dynamical-system type approach to 
robotic locomotion allows in addition to characterize 
the motion primitives in terms of self-organized 
attractors formed in the extended phase space of the 
robot and environment. Incorporating objects of the 
environment into the overarching dynamical system 
allows in consequence dynamical system approaches
also to classify computational models of affordance
\cite{zech2017computational}.
In this sense self-organizing attractors 
play an important role in the generation of useful 
behavior for the discovery of dynamic object 
affordances \cite{der2017selforganizing}.

%%%%%%%%%%%%%%%%%%%%%%%%%%%%%%%%%%%
\section*{Acknowledgments}
%%%%%%%%%%%%%%%%%%%%%%%%%%%%%%%%%%%
The support of the German Science Foundation
(DFG) is acknowledged.

%%%%%%%%%%%%%%%%%%%%%%%%%%%%%%%%%%%%%%%%%%%%%%%%%%%%%%%%%%%%%%%%%%%%%
%%%%%%%%%%%%%%%%%%%%%%%%%%%%%%%%%%%%%%%%%%%%%%%%%%%%%%%%%%%%%%%%%%%%%
%\bibliographystyle{splncs03}
%\bibliographystyle{unsrt}
%bibliography{trainCars}

%%%%%%%%%%%%%%%%%%%%%%%%%%%%%%%%%%%
%%%%%%%%%%%%%%%%%%%%%%%%%%%%%%%%%%%

\begin{thebibliography}{10}

\bibitem{tanaka2015control}
Motoyasu Tanaka and Kazuo Tanaka.
\newblock Control of a snake robot for ascending and descending steps.
\newblock {\em IEEE Transactions on Robotics}, 31(2):511--520, 2015.

\bibitem{granosik2014hypermobile}
Grzegorz Granosik.
\newblock Hypermobile robots--the survey.
\newblock {\em Journal of Intelligent \& Robotic Systems}, 75(1):147--169,
  2014.

\bibitem{hopkins2009survey}
James~K Hopkins, Brent~W Spranklin, and Satyandra~K Gupta.
\newblock A survey of snake-inspired robot designs.
\newblock {\em Bioinspiration \& biomimetics}, 4(2):021001, 2009.

\bibitem{pfotzer2017autonomous}
Lars Pfotzer, Sebastian Klemm, Arne R{\"o}nnau, Johann~Marius Z{\"o}llner, and
  R{\"u}diger Dillmann.
\newblock Autonomous navigation for reconfigurable snake-like robots in
  challenging, unknown environments.
\newblock {\em Robotics and Autonomous Systems}, 89:123--135, 2017.

\bibitem{sato2011applicability}
Takahide Sato, Takeshi Kano, and Akio Ishiguro.
\newblock On the applicability of the decentralized control mechanism extracted
  from the true slime mold: a robotic case study with a serpentine robot.
\newblock {\em Bioinspiration \& biomimetics}, 6(2):026006, 2011.

\bibitem{sato2011decentralized}
Takahide Sato, Takeshi Kano, and Akio Ishiguro.
\newblock A decentralized control scheme for an effective coordination of
  phasic and tonic control in a snake-like robot.
\newblock {\em Bioinspiration \& biomimetics}, 7(1):016005, 2011.

\bibitem{boyle2013adaptive}
Jordan~H Boyle, Sam Johnson, and Abbas~A Dehghani-Sanij.
\newblock Adaptive undulatory locomotion of a c. elegans inspired robot.
\newblock {\em IEEE/ASME Transactions on Mechatronics}, 18(2):439--448, 2013.

\bibitem{hirose2004biologically}
Shigeo Hirose and Makoto Mori.
\newblock Biologically inspired snake-like robots.
\newblock In {\em Robotics and Biomimetics, 2004. ROBIO 2004. IEEE
  International Conference on}, pages 1--7. IEEE, 2004.

\bibitem{liljeback2012review}
P{\aa}l Liljeb{\"a}ck, Kristin~Ytterstad Pettersen, {\O}yvind Stavdahl, and
  Jan~Tommy Gravdahl.
\newblock A review on modelling, implementation, and control of snake robots.
\newblock {\em Robotics and Autonomous Systems}, 60(1):29--40, 2012.

\bibitem{aoi2017adaptive}
Shinya Aoi, Poramate Manoonpong, Yuichi Ambe, Fumitoshi Matsuno, and Florentin
  W{\"o}rg{\"o}tter.
\newblock Adaptive control strategies for interlimb coordination in legged
  robots: a review.
\newblock {\em Frontiers in neurorobotics}, 11:39, 2017.

\bibitem{calisti2017fundamentals}
M~Calisti, G~Picardi, and C~Laschi.
\newblock Fundamentals of soft robot locomotion.
\newblock {\em Journal of The Royal Society Interface}, 14(130):20170101, 2017.

\bibitem{martin2016closed}
Laura Martin, Bulcs{\'{u}} S{\'{a}}ndor, and Claudius Gros.
\newblock {Closed-loop robots driven by short-term synaptic plasticity:
  Emergent explorative vs. limit-cycle locomotion}.
\newblock {\em Frontiers in Neurorobotics}, 10:12, 2016.

\bibitem{der2006rocking}
Ralf Der, Frank Hesse, and Georg Martius.
\newblock Rocking stamper and jumping snakes from a dynamical systems approach
  to artificial life.
\newblock {\em Adaptive Behavior}, 14(2):105--115, 2006.

\bibitem{sprowitz2013towards}
Alexander Spr{\"o}witz, Alexandre Tuleu, Massimo Vespignani, Mostafa
  Ajallooeian, Emilie Badri, and Auke~Jan Ijspeert.
\newblock Towards dynamic trot gait locomotion: Design, control, and
  experiments with cheetah-cub, a compliant quadruped robot.
\newblock {\em The International Journal of Robotics Research}, 32(8):932--950,
  2013.

\bibitem{van2009compliant}
Ronald Van~Ham, Thomas~G Sugar, Bram Vanderborght, Kevin~W Hollander, and Dirk
  Lefeber.
\newblock Compliant actuator designs.
\newblock {\em IEEE Robotics \& Automation Magazine}, 16(3), 2009.

\bibitem{calanca2016review}
Andrea Calanca, Riccardo Muradore, and Paolo Fiorini.
\newblock A review of algorithms for compliant control of stiff and
  fixed-compliance robots.
\newblock {\em IEEE/ASME Transactions on Mechatronics}, 21(2):613--624, 2016.

\bibitem{pfeifer2012challenges}
Rolf Pfeifer, Max Lungarella, and Fumiya Iida.
\newblock The challenges ahead for bio-inspired'soft'robotics.
\newblock {\em Communications of the ACM}, 55(11):76--87, 2012.

\bibitem{flash2005motor}
Tamar Flash and Binyamin Hochner.
\newblock Motor primitives in vertebrates and invertebrates.
\newblock {\em Current opinion in neurobiology}, 15(6):660--666, 2005.

\bibitem{ijspeert2002movement}
Auke~Jan Ijspeert, Jun Nakanishi, and Stefan Schaal.
\newblock Movement imitation with nonlinear dynamical systems in humanoid
  robots.
\newblock In {\em Robotics and Automation, 2002. Proceedings. ICRA'02. IEEE
  International Conference on}, volume~2, pages 1398--1403. IEEE, 2002.

\bibitem{khansari2011learning}
S~Mohammad Khansari-Zadeh and Aude Billard.
\newblock Learning stable nonlinear dynamical systems with gaussian mixture
  models.
\newblock {\em IEEE Transactions on Robotics}, 27(5):943--957, 2011.

\bibitem{paraschos2013probabilistic}
Alexandros Paraschos, Christian Daniel, Jan~R Peters, and Gerhard Neumann.
\newblock Probabilistic movement primitives.
\newblock In {\em Advances in neural information processing systems}, pages
  2616--2624, 2013.

\bibitem{amor2014interaction}
Heni~Ben Amor, Gerhard Neumann, Sanket Kamthe, Oliver Kroemer, and Jan Peters.
\newblock Interaction primitives for human-robot cooperation tasks.
\newblock In {\em Robotics and Automation (ICRA), 2014 IEEE International
  Conference on}, pages 2831--2837. IEEE, 2014.

\bibitem{park2017chaotic}
Jihoon Park, Hiroki Mori, Yuji Okuyama, and Minoru Asada.
\newblock Chaotic itinerancy within the coupled dynamics between a physical
  body and neural oscillator networks.
\newblock {\em PloS one}, 12(8):e0182518, 2017.

\bibitem{sandor2015sensorimotor}
Bulcs{\'u} S{\'a}ndor, Tim Jahn, Laura Martin, and Claudius Gros.
\newblock The sensorimotor loop as a dynamical system: How regular motion
  primitives may emerge from self-organized limit cycles.
\newblock {\em Frontiers in Robotics and AI}, 2:31, 2015.

\bibitem{chemero2007gibsonian}
Anthony Chemero and Michael~T Turvey.
\newblock {Gibsonian Affordances for Roboticists}.
\newblock {\em Adaptive Behavior}, 15(4):473--480, 2007.

\bibitem{nakhaeinia2011review}
Danial Nakhaeinia, SH~Tang, SB~Mohd Noor, and O~Motlagh.
\newblock A review of control architectures for autonomous navigation of mobile
  robots.
\newblock {\em International Journal of Physical Sciences}, 6(2):169--174,
  2011.

\bibitem{mamiya2018neural}
Akira Mamiya, Pralaksha Gurung, and John~C Tuthill.
\newblock Neural coding of leg proprioception in drosophila.
\newblock {\em Neuron}, 100(3):636--650, 2018.

\bibitem{kaplan2018sensorimotor}
Harris~S Kaplan, Annika~LA Nichols, and Manuel Zimmer.
\newblock Sensorimotor integration in caenorhabditis elegans: a reappraisal
  towards dynamic and distributed computations.
\newblock {\em Philosophical Transactions of the Royal Society B: Biological
  Sciences}, 373(1758):20170371, 2018.

\bibitem{sandor2018kick}
Bulcs{\'u} S{\'a}ndor, Michael Nowak, Tim Koglin, Laura Martin, and Claudius
  Gros.
\newblock Kick control: using the attracting states arising within the
  sensorimotor loop of self-organized robots as motor primitives.
\newblock {\em Frontiers in neurorobotics}, 12, 2018.

\bibitem{der2008predictive}
Ralf Der, Frank G{\"u}ttler, and Nihat Ay.
\newblock Predictive information and emergent cooperativity in a chain of
  mobile robots.
\newblock In {\em ALIFE}, pages 166--172, 2008.

\bibitem{ambe2018simple}
Yuichi Ambe, Shinya Aoi, Timo Nachstedt, Poramate Manoonpong, Florentin
  W{\"o}rg{\"o}tter, and Fumitoshi Matsuno.
\newblock Simple analytical model reveals the functional role of embodied
  sensorimotor interaction in hexapod gaits.
\newblock {\em PloS one}, 13(2):e0192469, 2018.

\bibitem{der2013behavior}
Ralf Der and Georg Martius.
\newblock Behavior as broken symmetry in embodied self-organizing robots.
\newblock In {\em Artificial Life Conference Proceedings 13}, pages 601--608.
  MIT Press, 2013.

\bibitem{der2012playful}
Ralf Der and Georg Martius.
\newblock {\em The Playful Machine: Theoretical Foundation and Practical
  Realization of Self-Organizing Robots}, volume~15.
\newblock Springer Science \& Business Media, 2012.

\bibitem{smith2005open}
Russell Smith.
\newblock Open dynamics engine, 2006.

\bibitem{gros2015complex}
Claudius Gros.
\newblock {\em Complex and adaptive dynamical systems: A primer}.
\newblock Springer, 2015.

\bibitem{wernecke2017test}
Hendrik Wernecke, Bulcs{\'u} S{\'a}ndor, and Claudius Gros.
\newblock How to test for partially predictable chaos.
\newblock {\em Scientific reports}, 7(1):1087, 2017.

\bibitem{zech2017computational}
Philipp Zech, Simon Haller, Safoura~Rezapour Lakani, Barry Ridge, Emre Ugur,
  and Justus Piater.
\newblock {Computational models of affordance in robotics: a taxonomy and
  systematic classification}.
\newblock {\em Adaptive Behavior}, 25(5):235--271, 2017.

\bibitem{der2017selforganizing}
Ralf Der and Georg Martius.
\newblock {Self-organized behavior generation for musculoskeletal robots}.
\newblock {\em Frontiers in Neurorobotics}, 11:8, 2017.

\end{thebibliography}
\end{document}